\documentclass[10pt,twoside,twocolumn]{IEEEtran}

\usepackage{graphicx,cite,amsmath,amssymb,hhline}

\newcommand{\ith}[1]    {{#1}^{\underline{\text{th}}}}

\newcommand{\PX}[1] {{\mathbb{P}}\left\{{#1}\right\}}

\newcommand{\EXs}[2] {{\mathbb{E}}_{{#1}}\!\!\left\{{#2}\right\}}

\newtheorem{thm}{Theorem}
\newtheorem{lem}{Lemma}

\newcommand{\ovgamma} {\overline{\gamma}}

\newcommand{\Pbc}[1] {P_{b\text{#1}}}

\newcommand{\Pe} {P_{e}}
\newcommand{\Pb} {P_{b}}

\newcommand{\Pex} {P_{e}^{\star}}
\newcommand{\Pbx} {P_{b}^{\star}}
\newcommand{\ovgx} {\ovgamma^{\star}}

\begin{document}

\markboth{IEEE Transactions on Information Theory, Vol. XX, No. Y, Month 2009}
        {Conti, Panchenko, Sidenko, \& Tralli: Log-Concavity Property of the Error Probability with Application \ldots}

\title{\mbox{}
Log-Concavity Property of the Error Probability with Application to Local Bounds for Wireless Communications
        }
\author{Andrea Conti, Dmitry Panchenko, Sergiy Sidenko, and Velio Tralli
\thanks{Andrea Conti and Velio Tralli: ENDIF at University of Ferrara and WiLAB c/o University of Bologna, Italy  (e-mail: {\tt a.conti@ieee.org}, {\tt vtralli@ing.unife.it}). Dmitry Panchenko: Department of Mathematics, 
Massachusetts Institute of Technology, Cambridge, MA and Texas A\&M University, College Station, TX, USA
 (e-mail: {\tt panchenk@math.tamu.edu}). Sergiy Sidenko: Department of Mathematics,
Massachusetts Institute of Technology, Cambridge, MA, USA
(e-mail: {\tt sidenko@math.mit.edu}).}
\thanks{This research was 
supported in part by the FP7 European project OPTIMIX (Grant Agreement 214625).}
}

\maketitle                   

\vskip -2cm
\begin{abstract}
A clear understanding the behavior of the error probability (EP) as a function of signal-to-noise ratio (SNR) and other
system parameters is fundamental for assessing the design of digital wireless communication systems.
We propose an analytical framework based on the log-concavity property of the EP which we prove
for a wide family of multidimensional modulation formats in the presence of Gaussian disturbances and fading.
Based on this property, we construct a class of local bounds for the EP that improve known generic bounds in a given
region of the SNR and are invertible, as well as easily tractable for further analysis. This concept is motivated by the fact
that communication systems often operate with performance in a certain region of interest (ROI) and, thus, it may be
advantageous to have tighter bounds within this region instead of generic bounds valid for all SNRs.
We present a possible application of these local bounds, but their relevance is beyond the example made in this paper.
\end{abstract}

\begin{keywords}
Error statistics, fading channels, local bounds, log-concavity, performance evaluation, probability.
\end{keywords}

\section{Introduction}

\PARstart{T}{he} performance evaluation for digital wireless communication systems in terms of bit error probability (BEP) and symbol error probability (SEP) requires a careful characterization of disturbances, such as noise and interference, as well as of the wireless channel impairments due to small-scale and large-scale fading (see, e.g., \cite{Jak:B95,Pro:B01,SimAlo:B04}). This can result in cumbersome expressions for the error probability (EP) which require numerical evaluation.\footnote{Hereafter, when EP is indicated without specification of BEP and SEP it means that the concept is valid for both BEP and SEP.}

At a first thought, this fact does not appear a relevant issue from the performance study point of view due to the increasing trend of computational power of computers. On the other hand, these cumbersome solutions do not provide a clear understanding of the performance sensitivity to system parameters, which is of great importance for system design, as well as they are often too complicated for further evaluation or  inversion, which is as example needed in order to obtain thresholds in adaptive communication systems (see, e.g., \cite{GolChu:97,ConWinChi:05,ConWinChi:07}). Moreover, it has to be emphasized that simple parametric approximations and bounds on the performance at lower layers, such as physical layer, can avoid long bit-level simulations in upper level protocols network simulators, provided that they are able to capture the main aspects affecting the performance at lower levels.

Mainly, but not only for these reasons, the derivation of approximations and bounds on the exact EP is still of interest in the communication theory community. An example is given by $M$-ary quadrature amplitude modulation ($M$-QAM), that is adopted in several standards for wireless communication systems, due to its bandwidth efficiency, and is largely studied in conjunction with adaptive techniques which change modulation parameters to maximize transmission rate for a given target BEP in wireless channels. In fact, although early work on $M$-QAM dates back to the early sixties
\cite{Cah:60,HanLuc:60,CamGla:62,HanLuc:62}, the evaluation of BEP
for arbitrary $M$ is still of current interest.\footnote{For a brief
history of $M$-QAM, see \cite{HanWebKel:B00}.} To briefly summarize some relevant results for additive white Gaussian noise (AWGN) channel we recall that: parameterized exponential approximations fitting simulative BEP are adopted in \cite{FosSal:83,QiuCha:99,HalHuHal:00,ChuGol:01}; approximations based on signal-space concepts were given in \cite{LuLetChuLio:99}; an exact method to derive the SEP was proposed in \cite{DonBeaWit:99}; a recursive algorithm exploiting the relationship among different constellation sizes was developed in \cite{YanHan:00}; exact expression of the BEP for general $M$ was derived in \cite{ChoYoo:02}. Comparisons among approximations, bounds and the exact solution in fading channels (with small-scale fading and large-scale fading, i.e., shadowing) are given in \cite{ConWinChi:05}, where it is shown that, for low and medium values of the signal-to-noise-ratio (SNR), approximations depart from exact solutions as the constellation size, $M$, increases. Moreover, small differences between exact solution and approximation in AWGN channel can become relevant when the instantaneous BEP is averaged over small-scale fading. Similarly, in systems with multichannel reception, known approximations depart from the exact EP as the diversity order increases \cite{SimAlo:B00,ConWinChi:03}.

It is well known that bounds carry more information than approximations and also enable system design based on the worst or best case.  Quite often bounds are tight to the exact EP only for high SNRs (namely asymptotic bounds). Here we are interested in deriving simple invertible bounds tight in a given  region of interest (ROI) for the EP (e.g., for the BEP of uncoded systems typical ROI's are $[10^{-3},10^{-1}]$ or $[10^{-4},10^{-2}]$).
 
In this paper, we define the concept of locally-valid bounds (in the following called {\em local bounds}), that are tight upper and lower bounds on the EP valid only within a given region of the EP and not for all SNRs. This concept is motivated by the fact that there is often a ROI for the performance of the system under consideration and it is  preferable to have tight bounds in this region instead of bounds valid for all SNRs which are far from the exact solution within the ROI. 

The behavior of the EP is important for the definition of local bounds. In fact, the proposed framework is based on its {\em log-concavity property}. We recall  that a function $z(\cdot)$ is log-concave if $\log z(\cdot)$ is concave.\footnote{In the paper notation $\log$ stands for natural logarithm.}
In most cases the EP is reported on log-scale and investigated as a function of the signal-to-total disturbance ratio in deciBel (dB). It is  commonly recognized that on this scale the function is concave in several cases of interest. Even though it is generally assumed, as the authors often acknowledge
, there is not known formal proof of the long-concavity of the EP (examples of related issues are: convexity properties in binary detection problems which were analyzed in \cite{Azi:96}, and some results for the asymptotic behavior of bounds that were investigated in \cite{LoyKosGag:07}). 

In this paper we introduce the problem of log-concavity for general multidimensional decision regions and we prove this property for a class of signals with constellation on a multidimensional regular grid in the presence of Gaussian distributed disturbances, such as thermal noise and interference. In fact, there are several wireless systems and situations in which the interference can be modeled as Gaussian distributed (see, e.g., \cite{McESta:84,Chi:98,ChiConAnd:99,GioChi:05,GioChiWin:05,LanSta:02}). After having proved the log-concavity in both AWGN and fading plus AWGN channels for single and multiple channels reception schemes, as examples of application, it will be shown how to take advantage of this property in order to simplify the derivation of bounds valid for all SNRs, as well as to define local upper bound (LUB) and local lower bound (LLB) valid in a given ROI. 
Moreover, the form of the local bounds and the fact that they are easily invertible also enables the derivation of local bounds for others relevant performance 
figures such as the EP-based outage probability, which is the probability that the EP averaged over small-scale fading exceed a given tolerable target value \cite{ConWinChiWint:L03}, also exploited for the evaluation of the mean spectral efficiency  for adaptive modulation techniques \cite{ConWinChi:07}. It is 
finally emphasized that the log-concavity property for the EP can have many others applications, thus its relevance
is beyond the applications illustrated in the paper.

The rest of the paper is organized as follows: in Sec. \ref{sec:LC} the log-concavity property of the EP is proved in AWGN and in AWGN plus fading for systems employing single and multiple channels reception, and in Sec. \ref{sec:LB} it is applied to define a new class of bounds and local bounds, with a discussion on possible applications. 
Finally, our conclusions are reported in Sec. \ref{sec:Conclusions}.

\begin{figure*}[!t]
\normalsize
\setcounter{equation}{8} %
\begin{eqnarray}
p\left(t,{\cal X}\right) 
&=&
1-\sum_{k=0}^{d}P_k \ \mu^{\otimes d}\bigl([-(a/2) e^t,\infty)^{d-k}\times[-(a/2) e^t,(a/2) e^t]^{k}
\bigr)
\nonumber
\\
&=&
1-\sum_{k=0}^{d}P_k \ \mu\bigl([-(a/2)e^t,\infty)\bigr)^{d-k} 
\mu\bigl([-(a/2)e^t,(a/2)e^t]\bigr)^k.
\label{ato2}
\end{eqnarray}
\hrulefill %
\end{figure*}

\section{Log-Concavity Property for the Error Probability}
\label{sec:LC}

In this section the log-concavity property for the EP  is discussed first for transmission in AWGN channel by highlighting the mathematical structure of the problem in the different applications of digital communications. Since we are interested in obtaining general results  
we base our framework on the origin of detection errors in the presence of Gaussian disturbances. Within this framework we will then prove the log-concavity property for the class of signals with constellation on a multidimensional regular grid (e.g., in the two-dimensional case this class includes the well known $M$-QAM constellation).
Finally, we will address the log-concavity property in systems with AWGN plus fading channels.

Consider a set $\cal X$ of $M$ constellation points on a $d$-dimensional signal space, i.e.  
$
{\cal X} = \Bigl\{x_i\in \mathbb{R}^d : i=1,\ldots, M \Bigr\}.
$
 Let us consider an arbitrary probability distribution on the set $\cal X$
and let $p_i$ denote the probability of a point $x_i\in {\cal X}$ for $1\leq i\leq M$
(we arbitrarily order points in $\cal X$ without loss in generality).
For each $x_i \in {\cal X}$ let us define its neighborhood
\setcounter{equation}{0} %
\begin{equation}
R_i = \Bigl\{x\in \mathbb{R}^d : |x-x_i| = \min_{j} |x-x_j|\Bigr\}
\end{equation}
as the set of points closest to $x_i$ in $\cal X$.
Suppose that we transmit a point $x_i$ with probability $p_i$ in an AWGN channel, hence we receive $x_i+\sigma g$ where $\sigma>0$, and $g$ has
a standard Gaussian distribution ${\cal N}(0,I)$ on $\mathbb{R}^d$ with mean zero and 
identity covariance matrix.\footnote{Note that $\sigma^2$ is inversely proportional to the SNR.} We classify
each point according to a region $R_i$ that it belongs to, which means that we make
an error if $x_i+\sigma g\not\in R_i$ or 
\begin{equation}
g\not\in\sigma^{-1}(R_i-x_i)=\bigl\{
\sigma^{-1}(y-x_i) : y\in R_i
\bigr\} \,.
\end{equation}
Through the change of variable\footnote{Which is strictly related to the transformation of the SNR from the linear to the dB scale.} $\sigma^{-1}=e^t$, the total probability of making an error results in \footnote{Notation $\PX{{\cal{A}}}$ stands for probability of event ${\cal{A}}$.}
\begin{eqnarray}
p\left(t,{\cal X}\right)
&=&
\sum_i p_i\, \PX{g\not\in e^{t}(R_i-x_i)} \nonumber \\
&=&
1- \sum_i p_i\, \PX{g\in e^t (R_i-x_i)}. 
\end{eqnarray}
If we denote by $\mu$ a standard Gaussian measure on $\mathbb{R}$ then the distribution of 
vector $g$ is a product measure $\mu^{\otimes d}$ and, therefore,
\begin{equation}
p\left(t,{\cal X}\right)
=\sum_ip_i\, \mu^{\otimes d}\left(e^t \overline{R_i'}\right)=
1-\sum_ip_i\, \mu^{\otimes d}\left(e^t R_i'\right),
\label{PE}
\end{equation}
where we denoted by $R_i'=R_i-x_i$ the region $R_i$ translated by $x_i$, and by $\overline{R_i'}$ the region $\mathbb{R}^d-R_i'$.

The function $p\left(t,{\cal X}\right)$ is the error probability in the detection of digital signals, either coded or uncoded, as a function of signal-to-noise ratio $t$ (in logarithmic scale). Proving the log-concavity property of this function is a challenging task. In fact, the log-concavity of single terms in the linear combination of eq. (\ref{PE}) depends on the specific structure of regions ${R_i'}$ and $\overline{R_i'}$ and in any case a possible linear combination of log-concave functions is not necessarily log-concave.


Only in few special cases, as example when all the regions $R_i'$ have the same measure and a special symmetry around the axis intersecting points $0$ and $-x_i$, these issues may be  overcome with the help of the Prekopa-Leindler theorems \cite{Pre:73,Lei:72} which states that {\it the function $F({\bf x})=\int_A f({\bf x},{\bf y}) d{\bf y} $, where ${\bf x}\in \mathbb{R}^n,{\bf y}\in \mathbb{R}^m $, is log-concave in $\mathbb{R}^n$ if $f({\bf x},{\bf y}) $ is log-concave in $\mathbb{R}^{n+m}$ and $A$ is a convex subset of $\mathbb{R}^m$}.

Two examples, one for uncoded system and the other for coded system,  will illustrate these simple cases below. 
 In all the other cases an inspection of log-concavity property should be based on the specific properties of the signal set ${\cal X}$.
In the next Sec. \ref{sec:lc-grid-awgn} we will provide the proof of log-concavity property for the specific case of signals with constellation on a multidimensional regular grid, which covers all the relevant applications based on $M$-QAM signaling.

\begin{figure}[!t]
\centerline{\includegraphics[width=1.1\linewidth,draft=false]{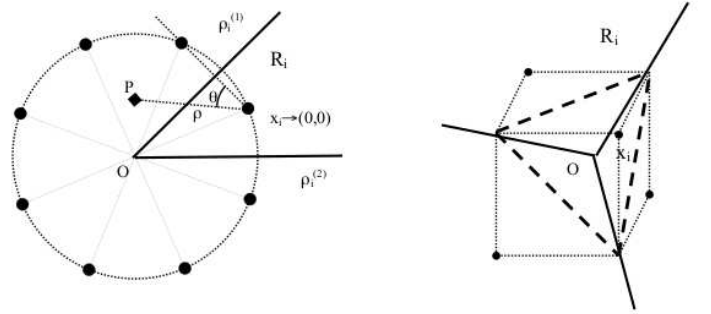}}
\caption{Decision regions $R_i$ for the examples 1 (left) and 2 (right) of Sec. \ref{sec:LC}, respectively.}
\label{fig:region}
\end{figure}

{\it Example 1 ($M$-PSK):} Let us consider a 2-dimensional signal set ${\cal X}$ where the $M$ points are regularly placed on a circle. 
The angular separation between closest points is $2\pi/M$ (see Fig. \ref{fig:region}-left). This is the signal set used by $M$-PSK signaling.
The regions $R_i'$ are circular sectors, have the same form and the same measure, and are convex. The same holds for regions $\overline{R_i'}$, which are concave instead. If we split each region $\overline{R_i'}$ in two parts, $\rho^{(1)}_i$ and $ \rho^{(2)}_i$, using the line connecting the points (0,0) and $-x_i$, all these sub-regions have the same Gaussian measure and are convex. Since $p\left(t,{\cal X}\right)
=2 \mu^{\otimes 2}\left(e^t \rho^{(1)}_i\right)$, the log-concavity of a single term has to be checked. The 2-dimensional Gaussian measure can be evaluated by using polar coordinates\footnote{The origin is the point $x_i$ and $\theta$ is the angle with respect to the line orthogonal to $\rho^{(1)}_i$ boundary.} in $\mathbb{R}^2$ as
\begin{equation}
 \mu^{\otimes 2}\left(e^t \rho^{(1)}_i\right)=\frac{1}{2\pi}\int_{-\pi/2 +\pi /M}^{\pi/2} e^{-S^2(\theta)e^t} d\theta
\end{equation}
where $ S(\theta)= \sin(\pi /M)/cos(\theta)$  describes the boundary of region $\rho^{(1)}_i$. Since the function $e^{-s^2(\theta)e^t}$ is log-concave for $(t,\theta )\in \mathbb{R}\times [-\pi/2 +\pi /M,\pi/2] $, the 
 Prekopa-Leindler theorem\footnote{Here, the domain is 
restricted to $\mathbb{R} \times [-\pi /2,\pi/2]$.} assures that $p\left(t,{\cal X}\right)$ is log-concave with respect to $t$.

{\it Example 2 (Parity check linear block codes and BPSK):} Let us consider the $k+1$-dimensional signal set ${\cal X}$  representing signals obtained by combining a simple parity-check $(k+1,k)$ binary block code and binary antipodal modulation. All the $M=2^k$ points are placed on (half of) the vertexes of a $k+1$-dimensional cube and are equidistant from the origin. Each point has $n_c=k(k+1)/2$ closest points or neighbors and each region $R_i$ is bounded by $n_c$ faces in the $k+1$-dimensional space (see Fig. \ref{fig:region}-right). All regions $R_i'$ have the same form and the same measure, and are convex. The same holds for regions $\overline{R_i'}$, which are concave instead.
Let us now simplify the example to $k=2$ for better understanding. We have 4 equidistant points placed on 4 vertices of a cube. Regions $R_i'$ and $\overline{R_i'}$ are bounded by $nc=3$ planes intersecting in $-x_i$. If we split each region $\overline{R_i'}$ into three parts, $\rho^{(1)}_i$, $ \rho^{(2)}_i$ and $\rho^{(3)}_i$, using three half-planes generated by the line connecting the points (0,0,0) and $-x_i$, all these sub-regions have the same Gaussian measure and are convex. Since $p\left(t,{\cal X}\right)
=3 \mu^{\otimes 3}\left(e^t \rho^{(1)}_i\right)$, the log-concavity of a single term has to be checked. By using cylindrical coordinates\footnote{Here, the origin is the point $x_i$, $z$ is the coordinate along the  line orthogonal to $\rho^{(1)}_i$ boundary, $\theta$ is the angle on the plane orthogonal to $z$-axis.} in $\mathbb{R}^3$, the 3-dimensional Gaussian measure can be evaluated as
\begin{equation}
 \mu^{\otimes 3}\left( e^t \rho^{(1)}_i\right) =\frac{1}{\pi}\int_{-\pi/2 +\beta}^{\pi/2}  \! Q\left( \sqrt{(1+S^2(\theta ))e^t4/3}\right) d\theta
\end{equation}
where $z S(\theta)= z/(\sqrt{3}cos(\theta))$ with $z\ge \sqrt{2/3}$ describes the boundary of region $\rho^{(1)}_i$, $\beta=\arcsin(\sqrt{2/3})$, and $Q(.)$ is the Gaussian Q-function.
Since the function $Q\left( \sqrt{(1+S^2(\theta ))e^t4/3}\right)$ is log-concave\footnote{Note that $Q(\sqrt{x})$ is log-
concave, whereas $(1+S^2(\theta))e^t$ is convex.} for $(t,\theta )\in \mathbb{R}\times [-\pi/2 +\arcsin(\sqrt{2/3}),\pi/2] $, the 
 Prekopa-Leindler theorem assures that $p\left(t,{\cal X}\right)$ is log-concave with respect to $t$.

\begin{figure*}[!th]
\normalsize
\setcounter{equation}{20} %
\begin{eqnarray}\label{eq:exactinstBEP}
\Pb(e|\gamma)=
\frac{2}{\sqrt{M}\log_2 M 
}
\sum_{k=1}^{\log_2 \sqrt{M} \ \ }
\sum_{i=0}^{(1-2^{-k})\sqrt{M}-1} 
(-1)^{\left\lfloor\frac{i \cdot 2^{k-1}}{\sqrt{M}}\right\rfloor}
\left( 2^{k-1} - \left\lfloor\frac{i \cdot 2^{k-1}}{\sqrt{M}} +
\frac{1}{2}\right\rfloor \right) 
\  \text{erfc}\left[ (2i+1) \sqrt{\frac{3 
\gamma}{2(M-1)}}\ \right] 
\end{eqnarray}
\hrulefill %
\end{figure*}

\subsection{Log-Concavity property: signals with constellation on a multidimensional grid}
\label{sec:lc-grid-awgn}

Given $a>0,$ consider a set $\cal X$ of $M=n^d$ points on $\mathbb{R}^d$ 
\setcounter{equation}{6} 
\begin{equation}
{\cal X} = \Bigl\{(k_1 a,\ldots, k_d a) : k_l = 1,\ldots, n \mbox{ for all } l\leq d\Bigr\}
\end{equation}
that form a regular finite grid on $\mathbb{R}^d$ with each coordinate taking $n$ possible values
$a,2a, \ldots, na$.\footnote{Without loss of generality ${\cal X}$ can be translated.}
Since $\cal X$ is a regular grid, all sets $R_i'$ take a particularly simple form, namely,
each such set is equal to one of the $d+1$ sets given by
\begin{equation}
[-a/2,\infty)^{d-k}\times[-a/2,a/2]^{k}
\,\,\mbox{ for }\,\,
k=0,\ldots,d 
\label{sets}
\end{equation} 
up to a permutation of coordinates. The product measure $\mu^{\otimes d}$ is invariant under
permutation of coordinates,  thus we can identify each set $R_i'$ with one of
the sets in (\ref{sets}). 
If $P_k$ is the sum of probabilities $p_i$ of points contained in the regions of type $k$,
then we obtain \eqref{ato2}.

We note that making a change of variables $t\to \left(t-\log (a/2)\right)$
it suffices to consider the case of $a/2=1.$
Let us now define
\setcounter{equation}{9}
\begin{subequations}\label{eq:gh}
\begin{eqnarray}
g&:=&g(e^t)\triangleq \mu\left(\left[-e^t,\infty\right)\right)\,,\\
h&:=&h(e^t)\triangleq \mu\left(\left[-e^t,e^t\right]\right)\,.
\end{eqnarray}
\end{subequations}
This leads to the following representation
\begin{equation}
p\left(t,{\cal X}\right)=1-\sum_{k=0}^d P_kg^{d-k}h^k\,.
\end{equation}
Since $g=(h+1)/2$ we obtain
\begin{equation}
p\left(t,{\cal X}\right)=1-H\left(h\left(e^t\right)\right)=p\left(n,d,t\right)\,,
\end{equation}
where
\begin{equation}\label{eq:def_H(h)}
H\left(h\right)\triangleq \sum_{k=0}^d P_k\left(\frac{h+1}{2}\right)^{d-k}h^k
=\sum_{k=0}^d P_k H_k(h)\,.
\end{equation}

\noindent {\it Remark.} Notice that all derivatives of $H$ with respect to $h$  are nonnegative
and $H(1)=1.$ 

We now prove the log-concavity of $p\left(n,d,t\right)$ with respect to $t$ starting with 2 Lemmas. The main Theorem with proof will follow.
\begin{lem}\label{lem:inequality_H(h)}
For any $h\in [0,1]$ and with $H(h)$ given by \eqref{eq:def_H(h)} the following inequality holds
\begin{equation}\label{eq:inequality_H(h)}
\left(1-h\right)H'\left(h\right)^2
-\left(1-H\left(h\right)\right)\left(H'\left(h\right)-\left(1-h\right)H''\left(h\right)\right) \ge 0 \,.
\end{equation}
\end{lem}

\begin{lem}\label{lem:non-negativity}
If $d>1$, for any $h\in [0,1]$ and with $H_{k}(h)$ given by \eqref{eq:def_H(h)} the following inequality holds 
\begin{eqnarray}\label{eq:inequality_H_km(h)}
&&\left(1-h\right)H_k'(h) H_m'(h)\nonumber \\ &-&\left[ 1-H_k(h)\right] 
 \left[ H_m'(h)-\left(1-h\right)H_m''(h)\right] \ge 0 \,.
\end{eqnarray}
\end{lem}

\noindent We will prove this Lemma in the Appendix and now
show how Lemma 
\ref{lem:inequality_H(h)}
implies the main Theorem \ref{thm:main}.

\vskip 0.5 truecm
\begin{thm}\label{thm:main}
For any $n,d\geq 1$ the function $t\to p\left(n,d,t\right)$ is log-concave.
\end{thm}
\vskip 0.5 truecm
\begin{proof} ({\it of Theorem \ref{thm:main}})
Let $G\left(e^t\right)=H\left(h\left(e^t\right)\right)$.
A simple calculation gives
\begin{eqnarray}
&&\frac{d^2}{dt^2}\log\left(1-G\left(e^t\right)\right)
 =  -e^t \\ 
 &\times& \frac{\left(1-G\left(e^t\right)\right)G'\left(e^t\right)+e^tG'\left(e^t\right)^2+e^t\left(1-G\left(e^t\right)\right)G''\left(e^t\right)}{\left(1-G\left(e^t\right)\right)^2} \nonumber 
\end{eqnarray}
The right hand side is negative if an only if
\begin{equation}\label{eq:inequality_G(c)}
\left(1-G\left(c\right)\right)G'\left(c\right)+cG'\left(c\right)^2+c\left(1-G\left(c\right)\right)G''\left(c\right)\ge0\,,
\end{equation}
where $c=e^t>0$.
Since
\begin{eqnarray}
G'\left(c\right) &=&  H'\left(h\right)h'\left(c\right) \,, \nonumber \\
G''\left(c\right) &=& H''\left(h\right)h'\left(c\right)^2-H'\left(h\right)\cdot ch'\left(c\right) \,, \nonumber
\end{eqnarray}
and by definition of $h(c)$ giving
\begin{eqnarray}
h'\left(c\right) &=& \sqrt{\frac{2}{\pi}}e^{-\frac{c^2}{2}} \,, \nonumber \\
h''\left(c\right) &=&  
-c\ h'\left(c\right)\,, \nonumber
\end{eqnarray}
we can rewrite (\ref{eq:inequality_G(c)}) as
\begin{equation}
\left(1-H\right)\left(H'h'\left(1-c^2\right)+c\left(h'\right)^2H''\right)+c\left(h'\right)^2\left(H'\right)^2\ge0\,, \nonumber
\end{equation}
or, equivalently,
\begin{equation}
c\ h'\ge\left(c^2-1\right)\frac{\left(1-H\right)H'}{H''\left(1-H\right)+\left(H'\right)^2} \,.
\end{equation}
It is immediate to see that if $c\le1$ then the inequality holds. For $c>1$ the proof follows from
\begin{equation}
\frac{c\ h'}{c^2-1}\ge1-h\ge\frac{\left(1-H\right)H'}{H''\left(1-H\right)+\left(H'\right)^2}\,.
\end{equation}
Here, the left hand side inequality can be derived from
\begin{equation}\label{eq:inewL1}
\frac{c}{c^2-1}e^{-\frac{c^2}{2}}\ge\int_{c}^{\infty}e^{-\frac{x^2}{2}}dx\,, 
\end{equation}
which is verified
\footnote{
Both sides of \eqref{eq:inewL1} tend to $0$ as $c\to\infty,$ therefore, it is enough and simple  to show that for any $c>1$
$$
\frac{d}{dc}\left(\frac{c}{c^2-1}e^{-\frac{c^2}{2}}\right)\le\frac{d}{dc}\left(\int_{c}^{\infty}e^{-\frac{x^2}{2}}dx\right)\,.
$$
}
for $c>1$, 
and the right hand side inequality follows from Lemma 1.
\end{proof}

\vskip 0.5 truecm

{\it Remark. } 
The instantaneous BEP expression for coherent single reception $M$-QAM systems with arbitrary $M$ 
as a function of the instantaneous symbol SNR $\gamma$ is given by \eqref{eq:exactinstBEP} 
where $\lfloor x \rfloor$ denotes the integer part of $x$ \cite{ChoYoo:02}.  
One might consider to try to prove log-concavity directly using this explicit expression.
However, this seems to be a difficult task since sum of log-concave functions is not log-concave in general and the BEP is a linear combination of positive and negative terms containing the complementary error function\footnote{The complementary error function is in well known relationship with he Gaussian Q-function, i.e., $Q(x)=(1/2)\ \text{erfc}\left[x/\sqrt{2}\right]$.}, ${\text{erfc}}(\cdot)$,
making 
the analysis of \eqref{eq:exactinstBEP} not at all straightforward.

\vskip 0.5 truecm

\subsection{Log-Concavity property:  systems with AWGN plus Fading}

In the above proof of log-concavity of the function $t\to p(t, \cal X)$ 
the size of the grid $a>0$ was fixed. When we transmit a symbol related to the constellation point $x_i$ with fixed $a$ in AWGN plus fading channel, the receiver observes $F x_i + \sigma g$ where $F$ is a random variable (RV) representing the channel gain due to fading. This is equivalent to the observation of $x_i+\sigma g$ when the constellation has a random scaling parameter $a$ with the same statistics of the fading gain $F$.\footnote{E.g., this represents the case of flat fading channel and coherent reception.}
Thus, we show now that when the constellation $\cal X$ is scaled with a real random parameter $a$, the average of $p(t,a \cal X)$ over $a$ is still
log-concave in $t$. 

Let us now make the dependence of $p$ on $a$ explicit, through the change of variable $b=2\log a$, thus $a/\sigma=e^{t+b/2}$. We obtain the function $p(t+b/2, {\cal{X}})$, which is log-concave as a function of both variables
$(t,b)$ if $p(t, \cal X)$ is log-concave with respect to $t$. 
Hence, what follows is valid for all signal sets $\cal X$ with log-concave instantaneous EP function of $t$. To obtain the EP averaged over fading we have to evaluate the expected value of $p(t+b/2, \cal X)$ with respect to the RV $b$.

\vskip 0.5 truecm
\begin{thm}\label{thm:fading}
If $b$ has a log-concave PDF then the average of $p$ over $b$, that is $\overline{p}(t,{\cal{X}})=\EXs{b}{p(t+b/2,{\cal{X}})}$, is also log-concave.
\end{thm}
\vskip 0.5 truecm
\begin{proof} ({\it of Theorem \ref{thm:fading}})
Suppose that $b$ has a distribution with log-concave density,
that is density of the form $e^{-V(b)}$ for some convex function $V(b).$ Then\footnote{We omit here the dependence on set $\cal X$.}
$$
\overline{p}(t)=\int p(t+b/2) e^{-V(b)}db
$$
is the average of $p$ over $b.$ Since $p(t+b/2) e^{-V(b)}$ is log-concave
in both variables $(t,b),$ Prekopa-Leindler inequality \cite{Pre:73,Lei:72}
 implies that
$\overline{p}(t)$ is log-concave.
\end{proof}

Theorem \ref{thm:fading} shows that if the distribution of $b$ has
log-concave density then the average over $a$ is log-concave. This apply to several cases of interest (e.g., single and multiple channel reception in Rayleigh, Nakagami-$m$, and log-normal fading) thus leading to log-concave average EP. This can be verified by considering that if the PDF $f_{a^2}(\xi)$ of $a^2$ is given then the PDF of $b=\log a^2$ results
\setcounter{equation}{21} %
\begin{equation}
\label{eq:pdftransf}
f_b(z)=e^z\ f_{a^2}\left(e^z\right)\,.
\end{equation}
For Nakagami-$m$ fading (having $m\ge1/2$) the PDF of $a^2$ is given by\footnote{It is well known that Rayleigh fading is included in Nakagami-$m$ when $m=1$.}
\begin{equation}
f_{a^2}(\xi)=\frac{m^m}{\Gamma(m)} \xi^{m-1} e^{-m \xi}\,,
\end{equation}
from which by \eqref{eq:pdftransf} we obtain
\begin{equation}
f_{b}(z)=\frac{m^m}{\Gamma(m)}e^{m(z-e^z)}\,,
\end{equation}
that is log-concave in $z$ since $m(z-e^z)$ is concave and $m^m/\Gamma(m) >0$.
For log-normal fading (i.e., $a^2$ in dB is a zero-mean Gaussian RV with variance $\sigma_{\text{dB}}^2$) the PDF of $a^2$ is given by ($\nu=10/\log 10$)
\begin{equation}
f_{a^2}(\xi)=\frac{\nu}{\sqrt{2\pi}\sigma_{\text{dB}}\xi} e^{-\frac{(10\log_{10} \xi)^2}{2 \sigma_{\text{dB}}^2}}\,,
\end{equation}
from which by \eqref{eq:pdftransf} we obtain
\begin{equation}
f_{b}(z)=\frac{\nu}{\sqrt{2\pi} \sigma_{\text{dB}}} e^{-\frac{\nu^2 }{2 \sigma_{\text{dB}}^2}z^2} \,,
\end{equation}
that is log-concave in $z$ since $ \sigma_{\text{dB}}$ is positive .
For maximal ratio combining of $N$-branches i.i.d. Rayleigh fading, the PDF of $a^2$ at the combiner output  is given by
\begin{equation}
f_{a^2}(\xi)=\frac{1}{(N-1)!} \xi^{N-1} e^{- \xi}\,,
\end{equation}
from which by \eqref{eq:pdftransf} we obtain
\begin{equation}
f_{b}(z)=\frac{1}{(N-1)!}e^{Nz-e^z}\,,
\end{equation}
that is log-concave in $z$ since $N$ is positive and $Nz-e^z$ is concave.

It is also important to remark that the log-concavity property for the EP proved above can have several applications, thus its relevance is beyond what illustrated in the following sections where an application example for bounds and local bounds is provided.

\section{Bounds and Local Bounds on Log-Concave Error Probability}
\label{sec:LB}
In this section it is shown how to take advantage of the log-concavity property.
for the derivation of bounds and local bounds which are analytically simple and invertible for further analysis. 
An example of application will be briefly discussed, addressing local bounds of relevant performance metrics for
adaptive $M$-QAM systems.
However, the application of the log-concavity is not limited to these cases (e.g., bounds for multidimensional modulations\footnote{See , e.g., \cite{BigEli:88, ChuMur:05}. The benefit provided by multidimensional constellations has been widely known in the design of coded modulation \cite{TraNgu:06,Wei:87}.} as well as for $M$-PSK can also be derived). 

\begin{figure}[!t]
\centerline{\includegraphics[width=0.9\linewidth,draft=false]{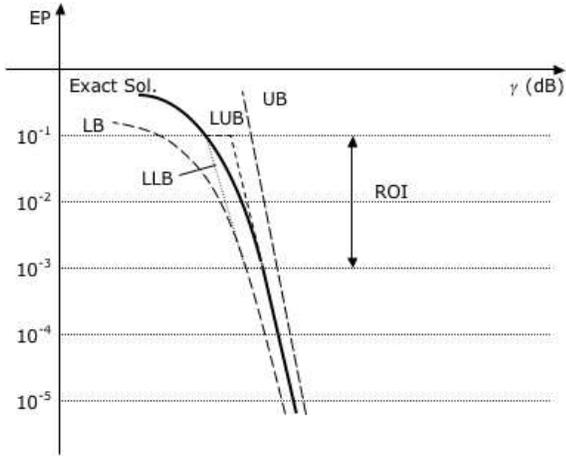}}
\caption{A general behavior for the EP in log-scale versus $\gamma$ in dB: concave exact solution, upper and lower bounds (UB, LB) as well as local upper and lower bounds (LUB, LLB) in the ROI of interest are reported.}
\label{fig:LB}
\end{figure}

The main idea to be exploited is that, due to the log-concave behavior proved in Sect. \ref{sec:LC}, the EP plotted in logarithmic scale versus the signal-to-total disturbance-ratio, $\gamma$, in deciBel (dB) is a concave function (see, e.g., Fig.\ref{fig:LB}).  After having identified the ROI, where the system typically operates, 
we aim to easily obtain tighter analytically tractable and invertible upper and lower bounds valid 
in the ROI.
The ROI is defined as the range $[P_{e\text{m}},P_{e\text{M}}]$ of the EP which is of interest in the specific application.

With the purpose to make a concrete example, in the following we consider the case of AWGN plus fading channel in which the performance is defined in terms of mean EP, the EP hereafter,  averaged over small-scale fading as a function of the mean $\gamma$, that is $\ovgamma$. Since the EP is monotonically decreasing in $\ovgamma$, the ROI corresponds to the SNR range $[\ovgx_{\text{M}},\ovgx_{\text{m}}]$,  with $\Pe(\ovgx_{\text{m}})=P_{e\text{m}}$ and $\Pe(\ovgx_{\text{M}})=P_{e\text{M}}$.

Let us first consider bounds valid for all SNRs, that is for a ROI corresponding to SNR in the range $[0,\infty]$. This ROI includes asymptotic behavior of EP. 
As an example, it is worthwhile to recall that in several cases, such as in single and multiple channels reception fading channel with Rayleigh or Nakagami-$m$ PDF, the system
achieves a diversity $\mathcal{D}$ if the asymptotic error
probability is log-linear. This means that 
%
$P_{e}(\ovgamma)\approx v(\ovgamma) \triangleq {K}/{\ovgamma
^{\mathcal{D}}}$
%
 where $K$ is a constant depending on the asymptotic behavior.
In other words, a system with
diversity $\mathcal{D}$ is described by a curve of error
probability with a slope approaching $10/\mathcal{D}$ [dB/decade]
for large $\ovgamma$.

Thus, we focus our attention on systems with log-linear asymptotical mean EP. For these systems the log-concavity of the mean EP immediately implies that its asymptotic behavior provides an upper bound in the ROI $[0,\infty]$. 
Let us consider the usual EP, $\Pe(\ovgamma)$, and the asymptote 
$v(\ovgamma)$, both in logarithmic scale as a function of $\ovgamma$ (in dB).   
Note that on this scale the EP is concave whereas
$v(\ovgamma)$ is linear. 
It is clear that $\Pe(0)\leq
v(0)$, since the EP is less than or equal to $1/2$. Furthermore, since $\Pe(\ovgamma)$ and $v(\ovgamma)$ are both
decreasing, and the two curves approach at $\ovgamma \rightarrow
\infty$, then $\Pe(\ovgamma) \leq v(\ovgamma)$. Therefore, an upper bound to $\Pe(\ovgamma)$ can be easily defined as
\begin{equation}\label{eq:ub}
P_{e\text{UB}}(\ovgamma) \triangleq \min\left\{\frac{1}{2},v(\ovgamma)\right\}\,.
\end{equation}
The UB on the inverse EP, that is on the value of $\ovgamma$ required to reach a target EP, $\Pex < 1/2$, is thus given by
\begin{equation}
\ovgx_{\text{UB}}=\left(\frac{K}{\Pex}\right)^{1/\mathcal{D}}\,.
\end{equation}

To define local bounds let us now consider a generic ROI $[P_{e\text{m}},P_{e\text{M}}]$ with $0 < P_{e\text{m}} < P_{e\text{M}} < 1/2$. By shifting the asymptotic UB to touch the exact solution at extremes of the ROI, we define local bounds tighter than previously known bounds, easily invertible, and thus enabling further analysis. At the lower end of the ROI, that is for a target EP $\Pex=P_{e\text{m}}$, we can define $\Delta\ovgamma_{m}$(dB) as the difference, in dB, between the required $\ovgamma$ for the   asymptotic upper bound and the exact solution: 
\begin{equation}
\Delta\ovgamma_m({\text{dB}})=\ovgx_{\text{UB}} ({\text{dB}}) - \ovgx_{\text{m}} ({\text{dB}})\,.
\end{equation}
Then, in linear scale:
\begin{equation}
\Delta \ovgamma_m=\frac{\ovgx_{\text{UB}}(P_{e\text{m}})}{\ovgx_{\text{m}}}\,. 
\end{equation}
We now define the LUB in the ROI as:
\begin{equation}
\label{eq:LUB}
P_{e\text{LUB}}(\ovgamma) \triangleq \min\{P_{e\text{M}},P_{e\text{UB}}(\ovgamma \cdot \Delta\ovgamma_m)\}\,,
\end{equation}
which is an invertible upper bound within the ROI. In fact, for a target EP $\Pex$ in the ROI, the LUB on the required $\ovgamma$ becomes:
\begin{equation}
\label{eq:LUBSNR}
\ovgx_{\text{LUB}}=\frac{\ovgx_{\text{UB}}(\Pex)}{\Delta \ovgamma_m}=\ovgx_{\text{m}} \left(\frac{P_{e\text{m}}}{\Pex}\right)^{1/\mathcal{D}} \,.
\end{equation}
Thus, to define the LUB one needs to know the exact required SNR at one point, namely, at the lower end of the ROI.

Similarly, one can define the invertible LLB, which is a lower bound within the ROI, by shifting the UB of $\Delta \ovgamma_M$ referred to $P_{e\text{M}}$. This needs  only the knowledge of the required SNR for the EP at the upper end of the ROI, $\ovgx_{\text{M}}$. The LLB is given by:
\begin{equation}
\label{eq:LLB}
P_{e\text{LLB}}(\ovgamma) \triangleq \max\{P_{e\text{m}},P_{e\text{UB}}(\ovgamma \cdot \Delta\ovgamma_M)\}\,.
\end{equation}
The LLB on the required $\ovgamma$ results in
\begin{equation}
\label{eq:LLBSNR}
\ovgx_{\text{LLB}}=\frac{\ovgx_{\text{UB}}(\Pex)}{\Delta \ovgamma_M}=\ovgx_{\text{M}} \left(\frac{P_{e\text{M}}}{\Pex}\right)^{1/\mathcal{D}}\,.
\end{equation}

A qualitative example of bounds and local bounds within the ROI is reported in Fig.\ref{fig:LB}. At this point it is important to emphasize that, while the LUB is, within the ROI, a tighter bound than the asymptotic UB and still invertible, the LLB obtained by translation of the UB can be worse compared to known LB, but in the other hand, the LLB is easily invertible enabling further analysis.\footnote{Asymptotic expressions for the EP in the form of $v(\ovgamma)$ can be found using ``systematic" approaches when exact EP expressions
are not available or asymptotic expressions can not be easily deduced from well-know (but often complicated) expressions (see, e.g., \cite{WanGia:03,GhaPas:95}).}

{\it Remark:} the log-concavity property opens the way for the definition of other classes of bounds, such as based on tangent in the extremes of the ROI or based on saddle-point (steepest descent method). Local bounds here proposed have the advantage of being simple and analytically tractable for further analysis.

We now discuss a possible application of local bounds on direct and inverse BEP to the evaluation of relevant performance metrics for adaptive $M$-QAM systems. Let us consider as an example, $M$-QAM with coherent detection and $N$-branches MRC,  whose exact BEP averaged over i.i.d. Rayleigh fading is given 
in \cite{ConWinChi:07} and its asymptotic upper bound is in the form $\Pbc{UB}\left(\ovgamma\right) \triangleq \min\left\{ 1/2, T_N(M)/ \ovgamma^N  \right\}$ where $T_N(M)$ depends only on the constellation-size and the diversity order.
From $\Pbc{UB}\left(\ovgamma\right)$ one can obtain the upper bound on the inverse BEP, which is the bound on the SNR required to achieve a target BEP equal to $\Pbx<1/2$, and from this invertible LUB and LLB through \eqref{eq:LUBSNR} and \eqref{eq:LLBSNR}, respectively. This enables the derivation of LLB and LUB on the error outage (EO), outage
probability based on the EP \cite{ConWinChiWint:L03,ConWinChi:J03}, which is an
appropriate QoS measure for digital mobile radio when small-scale fading is superimposed on shadowing (typically modeled with log-normal distribution \cite{Jak:B95,ErcEtal:99}).\footnote{The EO becomes the bit EO (BEO) or the symbol EO (SEO) when respectively related to the BEP or the SEP.} 
In systems with slow adaptive modulation,\footnote{What follows is also valid for fast adaptive modulation for which instantaneous EP and SNR are considered instead of those averaged over small-scale fading \cite{GolChu:97,ConWinChi:07}.} for a given target BEP, $\Pbx$, the spectral efficiency (SE) is a discrete RV\ with
distribution that depends on the SNR thresholds and on how they are computed (i.e., on the
BEP expression of the given system configuration). 
Let $M_{j}$ and $\ovgx_{\text{ dB,}j}$ be the $\ith{j}$ element
from the set of possible constellation sizes and corresponding SNR
threshold (in dB), respectively, to achieve a target BEP. Then, the mean SE results in
\begin{eqnarray}
\label{eq:NT} \eta &=& 
\sum_{j=0}^{J-1}  \tilde{M}_{j} \left[F_{\ovgamma_{\text{dB}}}(\ovgx_{\text{dB,}j+1}) -
F_{\ovgamma_{\text{dB}}}(\ovgx_{\text{dB,}j})\right] \nonumber \\
&+&
\tilde{M}_{J} \left[1 - F_{\ovgamma_{\text{dB}}}(\ovgx_{\text{dB,}J})
\right] \,,
\end{eqnarray}
where $\tilde{M}_{k}=\log_2 M_{k}$ and $F_{\ovgamma_{\text{dB}}}(
\cdot )$ is the cumulative distribution function (CDF) of
$\ovgamma_{\text{dB}}=10 \log_{10} \ovgamma$.
By substituting in \eqref{eq:NT} the required SNRs, $\ovgx_{\text{dB},j}$ with LUB, $\ovgx_{\text{LUB\ dB},j}$, we obtain a LLB on the mean SE allowing a conservative design of the communication system with different constellation-sizes and diversity orders.

\section{Conclusions}
\label{sec:Conclusions}

In this work we proved an important property of the error probability as a function of signal-to-noise-ratio in dB for AWGN channel as well as AWGN plus fading channels with single and multiple channels reception. In particular, we proved that the error probability is log-concave for a wide class of multidimensional modulation formats which include $M$-QAM for two dimensions. This property can have several applications. As example, we exploited log-concavity to derive upper and lower bounds and to define local bounds that are tight in a given region of interest for the error probability. We also discussed an application of local bounds highlighting the possibility of easy computation for the inverse of EP formulas without loosing significant accuracy in the evaluation of figures of merit interesting in wireless communications. However, we believe that the relevance of log-concavity property goes beyond the example provided in the paper and may be exploited for other different purposes.

\section*{Acknowledgments}
The authors would like to thank the Editor and the anonymous Reviewers for their suggestions that helped the authors to improve the content of the paper. Authors would like also to thank M.~Chiani, M.~Win, and O.~Andrisano for helpful comments and discussions.


\begin{figure*}[!th]
\normalsize
\setcounter{equation}{48} %
\begin{eqnarray}\label{eq:rightL2}
&&
\sum_\alpha{d-m \choose \alpha}
\Bigl(2\left(h+h^2+\ldots+h^{m+\alpha}\right)-h^{m+\alpha}\left(hd+m\right)+\left(d-m-2\alpha\right)\Bigr) \nonumber
\\
&=&
\left(1-h\right)\sum_\alpha{d-m \choose \alpha}
\Bigl(
2h\left(1+2h+\ldots+\left(m+\alpha-1\right)h^{m+\alpha-2}\right)+h^{m+\alpha}d
\Bigr) 
\end{eqnarray}
\setcounter{equation}{49} %
\begin{eqnarray}\label{eq:r1}
r\left(1\right) & = & \sum_\alpha{d-m \choose \alpha}
\Bigl(d\left(m+\alpha-1\right)-2\left(1+2+\ldots+\left(m+\alpha-1\right)\right)\Bigr)
\nonumber \\
 & = & \sum_\alpha{d-m \choose \alpha}
 \Bigl(d\left(m+\alpha-1\right)-\left(m+\alpha-1\right)\left(m+\alpha\right)\Bigr)
 \nonumber \\
 & = & \sum_\alpha{d-m \choose \alpha}
 \Bigl(\left(-d-m+dm+m^2\right)+\alpha\left(d+2m-1\right)+\alpha^2\Bigr)
\end{eqnarray}
\hrulefill %
\end{figure*}

\section*{Appendix }

\begin{proof}[Proof of Lemma \ref{lem:inequality_H(h)}]
Inequality (\ref{eq:inequality_H(h)}) holds for $h=1$, hence, from now on we will assume that $0 \le h<1$.
First, let us prove this inequality in the one-dimensional case $d=1$. In this case
\begin{eqnarray}
H\left(h\right) &=& P_{0}\frac{h+1}2+P_{1}h \nonumber \\
H'\left(h\right) &=& \frac{1}{2}P_{0}+P_{1} \nonumber \\
H''\left(h\right) &=&  0\,. \nonumber
\end{eqnarray}
Therefore, we need to prove that $\left(1-h\right)H'\ge 1-H$ or
$$
\left(1-h\right)\left(\frac{1}{2}P_{0}+P_{1}\right)\ge P_{0}\frac{1-h}{2}+P_{1}\left(1-h\right),
$$
which results in the exact equality.
Let us now consider the case $d>1$. We write the factor $1-H$ in the second
term in (\ref{eq:inequality_H(h)}) as
\begin{eqnarray}
1-H\left(h\right)&=&\sum_kP_k\left(1-\left(\frac{h+1}{2}\right)^{d-k}h^k\right) \nonumber \\
&=&\sum_kP_k\left(1-H_k(h)\right) \nonumber
\end{eqnarray}
and let us think of the left hand side of (\ref{eq:inequality_H(h)}) as a
homogeneous quadratic form in $(P_k)_{0\leq k\leq d}$ of the type 
$$
\sum_{0\le k,m\le d}P_kP_mH_{k,m}\left(h\right)\ge0,
$$
where $H_{k,m}(h)$ is given by
\setcounter{equation}{37}
\begin{eqnarray}
H_{k,m}\left(h\right) & = & \left(1-h\right)H_k'(h) H_m'(h) \\ &-&\left[ 1-H_k(h)\right] \left[ H_m'(h)-\left(1-h\right)H_m''(h)\right] \nonumber
\,.
\label{eq:H_km}
\end{eqnarray}
Lemma \ref{lem:inequality_H(h)} then follows from Lemma \ref{lem:non-negativity}.
\end{proof}

\begin{proof}[Proof of Lemma \ref{lem:non-negativity}]
Let us start by recalling the following well known identities involving
binomial coefficients:
\begin{eqnarray}
\sum_{\alpha=0}^N{N \choose \alpha} & = & 2^N,\label{eq:sum_0}\\
\sum_{\alpha=0}^N\alpha{N \choose \alpha} & = & 2^{N-1}N,\label{eq:sum_1}\\
\sum_{\alpha=0}^N\alpha^2{N \choose \alpha} & = & 2^{N-2}N\left(N+1\right).\label{eq:sum_2}
\end{eqnarray}
As for notation, if $L$ is a linear combination of $(P_k)_{0\leq k\leq d}$  we denote with $\left\{ L\right\} _k$ the coefficient of $P_k$ in $L$. 
By definition of $H_{k,m}$ in \eqref{eq:H_km} 
to finish the proof, it is enough to show that for any $0\le k,m\le n$ 
\begin{equation}\label{eq:inequality_km}
\left(1-h\right)\left\{ H'\right\} _k\left\{ H'\right\} _m\ge\left\{ 1-H\right\} _m\left\{ H'-\left(1-h\right)H''\right\} _k.
\end{equation}
Since 
\begin{equation}\label{eq:H_k}
\left\{ H\right\} _k=g^{d-k}h^k=\left(\frac{h+1}{2}\right)^{d-k}h^k,
\end{equation}
we have
\begin{eqnarray}
&&
\left\{ H'\right\}_k =  \left(\left(\frac{h+1}{2}\right)^{d-k}h^k\right)'
=\left\{ H\right\} _k\frac{hd+k}{h\left(h+1\right)}\label{eq:H'_k}
\end{eqnarray}
and
\begin{eqnarray}
\left\{ H''\right\} _k & = & \left\{ H'\right\} _k\frac{hd+k}{h\left(h+1\right)}+\left\{ H\right\} _k\left(\frac{hd+k}{h\left(h+1\right)}\right)' \\
 & = & \left\{ H\right\} _k\frac{\left(k^2-k\right)+2hk\left(d-1\right)+h^2d\left(d-1\right)}{h^2\left(h+1\right)^2} \,. \label{eq:H''_k} \nonumber
 \end{eqnarray}
By plugging \eqref{eq:H_k}, \eqref{eq:H'_k} and \eqref{eq:H''_k} into
\eqref{eq:inequality_km} we obtain
\begin{eqnarray}
&&\frac{\left(1-h\right)\left\{ H\right\} _m\left(hd+m\right)}{1-\left\{ H\right\} _m}
\ge \nonumber \\ && \frac{\left(k-k^2\right)-2hk\left(d-1\right) + h^2\left(\left(2d-1\right)k+2d-d^2\right)+h^{3}d^2}{hd+k}  \nonumber 
\end{eqnarray}
The derivative of the right-hand side with respect to $k$ is equal to
\begin{equation}
\frac{h-1}{\left(hd+k\right)^2}\left(k^2+hd\left(2k-1\right)+h^2d\left(d-1\right)\right) \nonumber
\end{equation}
which is negative if $k>1/2$. Therefore, the right-hand side attains its maximum for $k=0$ or $k=1$.
The difference for $k=0$ and $k=1$ is
\begin{equation}
\frac{\left(d-1\right)\left(1-h\right)}{1+hd}h\geq 0 \nonumber
\end{equation}
and, thus, the maximum is attained at $k=0$ and we need to prove that
\begin{equation}
\frac{\left(1-h\right)\left\{ H\right\} _m\left(hd+m\right)}{1-\left\{ H\right\} _m}
\ge\frac{h^2\left(2d-d^2\right)+h^{3}d^2}{hd} \,. \nonumber
\end{equation}
This is equivalent to
\begin{eqnarray}
\label{eq:inequality_m}
&&hd\left(2^{d-m}-\left(h+1\right)^{d-m}h^m\right)
\ge \\ && 2h\frac{2^{d-m}-\left(h+1\right)^{d-m}h^m}{1-h} -\left(h+1\right)^{d-m}h^m\left(hd+m\right)\,. \nonumber
\end{eqnarray}
Using the fact that
\begin{equation}
\left(h+1\right)^{d-m}h^m=\sum_{\alpha=0}^{d-m}{d-m \choose \alpha}h^{m+\alpha} \nonumber
\end{equation}
and \eqref{eq:sum_0}, the left hand side of \eqref{eq:inequality_m}
can be rewritten as
\begin{eqnarray}\label{eq:leftL2}
&&hd\Bigl(\sum_\alpha{d-m \choose \alpha}-\sum_\alpha{d-m \choose \alpha}h^{m+\alpha}\Bigr) \nonumber \\
& = & hd\sum_\alpha{d-m \choose \alpha}\left(1-h^{m+\alpha}\right) \nonumber \\
&=& d\left(1-h\right)\sum_\alpha{d-m \choose \alpha}\left(h+h^2+\ldots+h^{m+\alpha}\right) 
\end{eqnarray}
Similarly, the right hand side of \eqref{eq:inequality_m} is equal to 
\begin{eqnarray}
&&2h\sum_\alpha{d-m \choose \alpha}\left(1+h+\ldots+h^{m+\alpha-1}\right) \nonumber \\ &-& \sum_\alpha{d-m \choose \alpha}h^{m+\alpha}\left(hd+m\right) 
=\sum_\alpha{d-m \choose \alpha} \nonumber \\ &\times& \left(2\left(h+h^2+\ldots+h^{m+\alpha}\right) - h^{m+\alpha}\left(hd+m\right)\right) \,. \nonumber
\end{eqnarray}
Using that by \eqref{eq:sum_0} and \eqref{eq:sum_1}
\begin{equation}\label{eq:sum_0_1}
\sum_\alpha{d-m \choose \alpha}\left(d-m-2\alpha\right)=0
\end{equation}
after some mathematical manipulations we obtain \eqref{eq:rightL2}.

Finally, comparing expansions for the left and right hand side, that is \eqref{eq:leftL2} and \eqref{eq:rightL2}, respectively, 
\eqref{eq:inequality_m} becomes
\begin{eqnarray*}
&&\sum_\alpha {d-m \choose \alpha}
 d\left(1+h+\ldots+h^{m+\alpha-2}\right) \\ &\ge&2 \sum_\alpha {d-m \choose \alpha} \left(1+2h+\ldots+\left(m+\alpha-1\right)h^{m+\alpha-2}\right) \,.
\end{eqnarray*}
Combining all the coefficients for each power of $h,$ the left hand side can be written as
$
r\left(h\right)=c_0+c_1h+\ldots+c_{d-2}h^{d-2}
$
where 
\begin{equation}
c_l = (d-2(l+1))\sum_{\alpha\geq (2+l-m)\vee 0}^{d-m}{d-m \choose \alpha} \,. \nonumber
\end{equation}
Notice that the sign of $c_l$ is determined by $d-2(l+1)$ so that
$c_l>0$ if $l\le l_0$ and $c_l\le0$ if $l>l_0$. Since $h\le1$, this gives
\begin{eqnarray*}
r\left(h\right) & = & c_0+c_1h+\ldots+c_{d-2}h^{d-2}\\
 & = & \left(c_0+c_1h+\ldots+c_{l_0}h^{l_0}\right) \\
 &+&\left(c_{l_0+1}h^{l_0+1}+\ldots+c_{d-2}h^{d-2}\right)\\
 & \ge & \left(c_0+c_1+\ldots+c_{l_0}\right)h^{l_0} \\ &+& h^{l_0+1}\left(c_{l_0+1}+\ldots+c_{d-2}\right)\\
 & \ge & h^{l_0}\left(c_0+c_1+\ldots+c_{d-2}\right)=h^lr\left(1\right) \,.
\end{eqnarray*}
It remains to show that $r\left(1\right)\geq 0$. We observe that $r\left(1\right)$ results in \eqref{eq:r1}.
Then, using \eqref{eq:sum_0}, \eqref{eq:sum_1}, and \eqref{eq:sum_2} we obtain
\begin{eqnarray*}
r\left(1\right) & = & 2^{d-m-2}\left[4\left(-d-m+dm+m^2\right) \right. \\ &+& \left. 2\left(d+2m-1\right)\left(d-m\right)+\left(d-m-1\right)\left(d-m\right)\right]\\
 & = & 3d^2-5d+3\left(d-1\right)m+dm+m^2
\end{eqnarray*}
which is positive for $d>1$.
\end{proof}

\bibliographystyle{IEEEtran}

\end{document}